\documentclass{article}
\usepackage{spconf,amsmath,graphicx}
\usepackage{amssymb}
\usepackage{amsfonts}
\usepackage{textcomp}
\usepackage{stfloats}
\usepackage{url}
\usepackage{verbatim}
\usepackage[utf8]{inputenc}
\usepackage[ruled,vlined,linesnumbered]{algorithm2e}
\usepackage{tikz}
\tikzstyle{arrow}=[->] 
\usetikzlibrary{calc,bayesnet}
\usepackage{mathtools}
\usepackage{multirow}
\usepackage{caption}
\usepackage{hyperref}
\hypersetup{%
  colorlinks=false,
  linkbordercolor=red,
  pdfborderstyle={/S/U/W 1}
}
\usepackage{subcaption}
\usepackage[makeroom]{cancel}
\usepackage{multirow}
\usepackage{booktabs}

\DeclareMathOperator*{\argmax}{arg\,max}
\newcommand{\shorteq}{%
    \scalebox{0.75}[1.0]{\ensuremath{=}}
}
\newcommand{\shortminus}{%
    \scalebox{0.75}[1.0]{\ensuremath{-}}
}

\title{RECURSIVE ESTIMATION OF USER INTENT FROM NONINVASIVE ELECTROENCEPHALOGRAPHY USING DISCRIMINATIVE MODELS
}

\name{
Niklas Smedemark-Margulies\textsuperscript{1}, 
Basak Celik\textsuperscript{2}, 
Tales Imbiriba\textsuperscript{2}, 
Aziz Kocanaogullari\textsuperscript{2}*,
Deniz Erdo{\u{g}}mu{\c{s}}\textsuperscript{2}
\address{
\textsuperscript{1} Khoury College of Computer Science, Northeastern University, Boston, MA, USA \\
\textsuperscript{2} Dept. of Electrical \& Computer Engineering, Northeastern University, Boston, MA, USA\\
}
\thanks{This work supported by NIH 2R01DC009834 as part of CAMBI and by NSF CBET-2117626.}
\thanks{*AK is currently at Analog Devices}
} 

\begin{document}
\ninept
\maketitle
\begin{abstract}
We study the problem of inferring user intent from noninvasive electroencephalography (EEG) to restore communication for people with severe speech and physical impairments (SSPI).
The focus of this work is improving the estimation of posterior symbol probabilities in a typing task. At each iteration of the typing procedure, a subset of symbols is chosen for the next query based on the current probability estimate. 
Evidence about the user's response is collected from event-related potentials (ERP) in order to update symbol probabilities, until one symbol exceeds a predefined confidence threshold. We provide a graphical model describing this task, and derive a recursive Bayesian update rule based on a discriminative probability over label vectors for each query, which we approximate using a neural network classifier. We evaluate the proposed method in a simulated typing task and show that it outperforms previous approaches based on generative modeling.
\end{abstract}


\section{Introduction}
\label{sec:intro}
Recent research has explored diverse approaches for estimating user intent from electroencephalography (EEG). Methods and applications vary greatly, from using invasive measurements and unprompted motor-imagery classification for typing~\cite{willett2021high}, to noninvasive query-and-response paradigms for wheelchair steering~\cite{pires2008visual}, and continuous frequency-domain measurement for video game control~\cite{van2013experiencing}, to name but a few.

Brain-computer interface (BCI) designs make use of many measurement paradigms, including motor imagery, error potentials, and visual stimulus responses~\cite{abiri2019comprehensive}. In particular, many methods rely on event-related potentials (ERPs) such as the P300~\cite{mak2011optimizing}. Ongoing work has explored modifications to these pre-existing approaches such as repeated presentation of stimuli~\cite{lin2018novel}, the fusion of evidence from multiple analysis paradigms~\cite{gonzalez2021feedback}, and the fusion of evidence from multiple sensors modalities~\cite{kalika2017fusion}. 

We focus on the domain of noninvasive brain-computer interfaces (BCIs) using a query-and-response paradigm called rapid serial visual presentation (RSVP)~\cite{acq10,orhan2012rsvp,dal2010online}. In this approach, a user seeks to select a target symbol from a predefined alphabet. The user is queried with a rapid sequence of candidate symbols from this alphabet while collecting continuous EEG data. 
The central task is to perform updates to symbol probabilities based on the user's measured response after these query stimuli.

Bayesian approaches have been widely adopted in building ERP-based communication systems. Prior work focused on EEG signal modeling includes techniques such as unsupervised learning algorithms~\cite{kindermans2012bayesian}, sparse models~\cite{zhang2015sparse}, and adaptive querying strategies~\cite{throckmorton2013bayesian}. 
Other work has focused on incorporating language model priors~\cite{mog15}.
These lines of research have also led to the creation of many open-source BCI frameworks incorporating these modeling strategies~\cite{memmott2021bcipy, HBM:HBM23730,10.3389/fnins.2013.00267}. 
Many of these Bayesian approaches naturally make use of generative models for computing likelihoods when updating symbol probabilities~\cite{orhan2013offline}. 
One challenge these methods face is that EEG datasets are often limited in size, with large variation between subjects, and significant noise and artifacts. 
This makes for an undesirable trade-off; training generative models on noisy, high-dimensional data is challenging, while reducing the data dimension in order to make techniques like covariance estimation feasible can also result in discarding useful information.

We avoid this trade-off by taking an alternative approach to performing Bayesian updates of symbol probabilities. Specifically, we derive a recursive update rule that only makes use of a discriminative probability distribution. In this way, we avoid the challenges of generative modeling, and can easily incorporate deep neural network models trained directly on the original high-dimensional input data. 

To evaluate the proposed model, we construct a simulated typing system using a large publicly-available RSVP benchmark dataset containing over 1 million stimulus presentations and corresponding EEG data~\cite{zhang2020benchmark}. We find that the proposed method outperforms baseline generative modeling strategies, both in balanced accuracy and in information transfer rate (ITR) during simulated typing. Our novel approach opens the door for future exploration using other classification techniques~\cite{lotte2018review}.

\section{Proposed Method}
\subsection{Problem Statement}
Consider a typing task in which a user seeks to type an unknown symbol $D$ from a fixed alphabet of size $A$. 
We begin with a prior probability distribution over the alphabet, based on previously typed symbols $\tau$.
At the $N^{th}$ iteration of the typing procedure, we use the current estimated probability of each symbol to select a set of $K$ symbols to query $q_N^{1}, \ldots q_N^{K}$; these are presented to the user in rapid succession.
We observe the $K$ EEG responses to the queried symbols $e_N^{1} \ldots e_N^{K}$, and compute a new estimated posterior probability for each symbol in the alphabet.
The focus of our work is to compare modeling strategies for computing the posterior probability over the alphabet.

In this work, we make two simplifying assumptions. First, we assume that the $K$ responses observed during a single query can be considered independently. This may be reasonable given that delay between neighboring symbols in a query is sufficiently large and the timing of 
event-related potentials in the user's EEG response is sufficiently well stereotyped to allow clear separation between neighboring responses. Second, we assume a uniform prior over the alphabet at the beginning of each attempted symbol, since our focus is only on the relative effects of different EEG signal modeling strategies.
\subsection{Probabilistic Graphical Model}
\begin{figure}[hb]
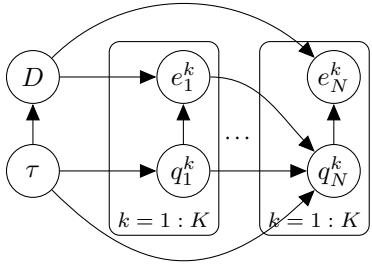

\centering
\tikz{
    \node[latent] (tau) at (0, 0) {$\tau$};
    \node[latent] (D) at (0, 1.25) {$D$};
    \node[latent] (q1k) at (2, 0) {$q_1^k$};
    \node[latent] (e1k) at (2, 1.25) {$e_1^k$};
    \plate {query1} {(e1k) (q1k)} {$k=1:K$};
    \node[latent] (qNk) at (4, 0) {$q_N^k$};
    \node[latent] (eNk) at (4, 1.25) {$e_N^k$};
    \plate {query2} {(eNk) (qNk)} {$k=1:K$};
    \node at ($(query1)!.5!(query2)$) {\ldots};
    \draw [style=arrow, out=90, in=270] (tau) to (D);
    \draw [style=arrow, out=0, in=180] (tau) to (q1k);
    \draw [style=arrow, out=90, in=-90] (q1k) to (e1k);
    \draw [style=arrow, out=90, in=-90] (qNk) to (eNk);
    \draw [style=arrow, out=-45, in=-135, looseness=1.3] (tau) to (qNk);
    \draw [style=arrow, out=0, in=180] (q1k) to (qNk);
    \draw [style=arrow, out=0, in=135] (e1k) to (qNk);
    \draw [style=arrow, out=0, in=180] (D) to (e1k);
    \draw [style=arrow, out=45, in=135] (D) to (eNk); 
}
\caption{Probabilistic graphical model for RSVP typing task after $N$ rounds of query and response, with $K$ symbols per query. $D$ - user's target symbol. $\tau$ - previously typed text. $q_N^k$ and $e_N^k$ - symbols and EEG responses during $N^{th}$ query.}
\label{fig:graphical-model}
\end{figure}

We seek an explicit formula for updating the estimated posterior probability of symbols in our alphabet.
We begin the typing procedure by observing previously typed symbols $\tau$. 
During the $N^{th}$ iteration of the query-and-response procedure, after presenting the $k^{th}$ symbol $q_N^k$ and observing the corresponding EEG response $e_N^k$, we are interested in the posterior distribution over symbols $D$; call this posterior $\pi_N^k(D)$.
In Fig.~\ref{fig:graphical-model}, we provide a probabilistic graphical model describing the relationship between these random variables.

For convenience, define
$Q$ as the set of all previously queried symbols $\{ q_{1:N-1}^{1:K} \} \cup \{ q_{N}^{1:k-1} \}$, 
define $q$ as the currently queried symbol $q_N^k$,
define $E$ as the corresponding set of EEG evidence $\{ e_{1:N-1}^{1:K} \} \cup \{ e_N^{1:k-1} \}$, and
define $e$ as the currently observed EEG evidence $e_N^k$.
Based on the graphical model, we can observe the following conditional independence relationships.
First, the target symbol is independent of the currently queried symbols given previously typed text and all previous queries and responses: $D \bot q_N^k | \tau, Q, E$. Second, the EEG response is independent of the previously typed text and all previous queries and responses, given the symbol currently presented and the true target symbol: $e \bot \tau, Q, E | D, q$. We apply these conditional independence relationships to derive a formula for updating our estimate of the posterior probability over symbols.

\subsection{Recursive Bayesian Update Rule}
\label{sec:recursive-update}

Here, we derive recursive update rules for both symbols that were presented and symbols not presented to the user, given the current query symbol $q$ and its respective response $e$.
Suppose that the current query symbol $q$ is $\alpha$; we can expand the posterior distribution for this symbol using Bayes' rule:
\begin{align}
    & \pi_N^k(D \shorteq \alpha) \coloneqq p( D \shorteq \alpha | \tau, Q, E, q \shorteq \alpha, e) \\
    & = \frac{
        p(e | D \shorteq \alpha, \tau, Q, E, q \shorteq \alpha) p(D | \tau, Q, E, q \shorteq \alpha)}{p(e | \tau, Q, E, q \shorteq \alpha)} \\
    & \propto p(e | D \shorteq \alpha, \cancel{\tau, Q, E}, q \shorteq \alpha) p(D \shorteq \alpha | \tau, Q, E, \cancel{q \shorteq \alpha}) \label{eqn:cancellations} \\
    & = p(e | D \shorteq \alpha, q \shorteq \alpha) p(D \shorteq \alpha | \tau, Q, E). \label{eqn:bayes-numerator}
\end{align}
The normalizing constant in Bayes' rule $p(e | \tau, Q, E, q \shorteq \alpha)$ does not depend on the target symbol $D$; thus we can ignore this term and normalize our distribution across the alphabet later. We apply the conditional indepedences previously observed to cancel terms in Eq.~\ref{eqn:cancellations}.
Note that $p(D \shorteq \alpha | \tau, Q, E)$ is exactly the posterior probability computed using all previous queries and evidence $\pi_N^{k-1}(D \shorteq \alpha)$.

Now, let $\ell \shorteq \{+, -\}$ be a discrete random variable representing the binary label of the currently queried symbol $q$, and $p(e,\ell)$ the joint distribution over responses and labels. We can use the Chapman-Kolmogorov equation and the product rule to isolate $e$ in the graph in Fig.~\eqref{fig:graphical-model}; that is, $p(e|D \shorteq \alpha, q\shorteq \alpha) = \sum_\ell p(e|\ell)p(\ell| D\shorteq \alpha, q\shorteq \alpha)$, where we used the conditional independence between $e$ and $D$ and $q$ given the label $\ell$.
Note that $\ell$ only depends on $D$ and $q$; $\ell$ will be positive when $D$ and $q$ match, and negative otherwise. Thus, we have
\begin{align}
    & \pi_N^k(D \shorteq \alpha) \propto p(e | D \shorteq \alpha, q \shorteq \alpha) \pi_N^{k-1}(D \shorteq \alpha) \label{eq:gen_update}\\
    & = \left[ \sum_\ell p(e | \ell) \underbrace{p(\ell | D \shorteq \alpha, q \shorteq \alpha)}_{p(\ell \shorteq +) = 1} \right] \pi_N^{k-1}(D \shorteq \alpha) \\
    & = p(e | \ell \shorteq +) \pi_N^{k-1}(D \shorteq \alpha) \\
    & = \frac{p(\ell \shorteq + | e) {p(e)}}{p(\ell \shorteq +)} \pi_N^{k-1}(D \shorteq \alpha) \\
    & \propto \frac{p(\ell \shorteq + | e)}{p(\ell \shorteq +)} \pi_N^{k-1}(D \shorteq \alpha)
    \coloneqq \gamma_N^k(D \shorteq \alpha). \label{eqn:unnormalized-update-seen}
\end{align}
Here, we define $\gamma_N^k(D \shorteq \alpha)$ as the unnormalized posterior probability for the current symbol $D \shorteq \alpha$.
Once again, we ignored a constant of proportionality $p(e)$, which does depend on the target symbol $D$. 

Now, we derive the update rule for symbols not presented to the user. For this, consider another candidate symbol $D \shorteq \beta \! \neq \! \alpha$. A nearly identical derivation can be followed to obtain an expression for $\gamma_N^k(D \shorteq \beta)$. The only difference is that, because this is not the currently presented symbol $q \shorteq \alpha$, the label variable $\ell$ that we introduce will have $p(\ell \shorteq -) = 1$, leading to:
\begin{align}
    \gamma_N^k(D \shorteq \beta) & = \frac{p(\ell \shorteq - | e)}{p(\ell \shorteq -)} \pi_N^{k-1}(D \shorteq \beta).
    \label{eqn:unnormalized-update-not-seen}
\end{align}
After applying these two rules to obtain unnormalized posteriors for all symbols, we can normalize across the alphabet to get our final posterior estimate for the presented query symbol $\alpha$ and all other symbols $\beta$. Let $\mathcal{D}$ represent the domain of our alphabet:
\begin{align}
    \pi_N^k(D \shorteq \alpha) = \frac{
        \gamma_N^k(D \shorteq \alpha)
    }{
        \sum\limits_{d \in \mathcal(D)} \gamma_N^k(d)
    }, \quad
    \pi_N^k(D \shorteq \beta) = \frac{
        \gamma_N^k(D \shorteq \beta)
    }{
        \sum\limits_{d \in \mathcal(D)} \gamma_N^k(d)
    }\, \cdot
    \label{eqn:normalized-update}
\end{align}

We have obtained an expression that, at each iteration, only requires two new quantities for each symbol: $p(\ell | e)$ and $p(\ell)$. The first term is a discriminative distribution over binary labels given EEG evidence, which we can approximate using classifier models such as neural networks. The second term is a prior distribution over labels. In this work, we simply approximate this with a uniform distribution over symbols in the alphabet. 

\section{Experimental Details}
In this section we describe our experimental results, which can be reproduced using our source code\footnote{\url{https://github.com/nik-sm/bci-disc-models/}}.

\subsection{Dataset and preprocessing} 
All experiments described here were performed using a large publicly-available RSVP benchmark dataset containing over $1$ million stimulus presentations~\cite{zhang2020benchmark}.

Data in this benchmark dataset were recorded at $1000$ Hz and made available at a down-sampled rate of $250$ Hz.
Each recording session contains two recording blocks of contiguous data.
We process each recording block as follows.
Two faulty channels are excluded as instructed by the dataset authors.
Data is filtered using a second-order infinite impulse response notch filter with central frequency $50$ Hz (regional AC line frequency) and quality factor of $30$, followed by a second-order Butterworth bandpass filter between $1$ and $20$ Hz. 
Data is then downsampled by $2$-fold, and finally segmented into partially overlapping trials. 
Each trial begins at the onset of a visual stimulus, and includes $500$ ms of subsequent data.
We provide Python code for easily downloading and preprocessing this dataset\footnote{\url{https://github.com/nik-sm/thu-rsvp-dataset}}.

We conduct all experiments using data pooled across subjects, and repeat each experiment using $5$ randomized train/test splits of the dataset.

\subsection{Discriminative Models}
We train deep neural network classifiers in order to make best use of the proposed discriminative framework. 
We present results from three neural network architectures. 
The first is based on EEGNet~\cite{lawhern2018eegnet}, a hand-designed 2D convolutional neural network (CNN). 
We modify its hyperparameters based on the short time duration of our pre-processed signal, and add an intermediate layer to produce a fixed latent dimension regardless of input dimensions.
The other two architectures are simple networks consisting of a repeated sequence of convolution, batch normalization~\cite{ioffe2015batch}, and non-linear activation, followed by a single linear layer. 
In one of these simple CNNs, we apply only 1D convolutions across the time axis; in the other, we apply only 2D convolutions across channels and time. See our code for full architecture details.

These models are coded in Python using PyTorch~\cite{paszke2019pytorch}, and trained for $25$ epochs using AdamW~\cite{loshchilov2017decoupled} optimizer with learning rate of $0.001$, decayed by a factor of $0.97$ after each epoch. 
Since our dataset has highly imbalanced classes (as is typical for RSVP tasks), models are trained to minimize weighted cross-entropy, with class weights set as the inverse of label fractions in the train set.
We evaluate downstream performance only using model parameters from the epoch of best balanced accuracy on a validation set consisting of a held-out $10\%$ of the train set.

In addition to neural network models, we also evaluate the proposed discriminative framework with a simple baseline classification pipeline consisting of per-channel z-scoring followed by logistic regression (LogR).

\subsection{Generative Models}
We compare the proposed approach against a more traditional generative modeling strategy. In this approach, we seek to perform the same Bayesian update, but try to directly model the likelihood term $p(e | D, q)$ from Eq.~\ref{eqn:bayes-numerator}. The challenge is that the distribution of EEG responses is high-dimensional, varies strongly between individuals, and suffers from significant measurement noise and artifacts. Therefore, we follow previous strategies~\cite{throckmorton2013bayesian, orhan2013offline, memmott2021bcipy} and adopt a multi-step process to greatly compress the data into one dimension, and finally use kernel density estimation (KDE) to compute the desired likelihood terms.

The compression stage of this pipeline consists of channel-wise z-scoring, concatenation of sensor channels, reduction using principal components analysis (PCA) to preserve $80\%$ percent variance, and finally feeding this reduced data into a classifier model to obtain an estimated log ratio of probabilities for the two classes. Next, we perform KDE using a Gaussian kernel with bandwidth $1$ to obtain a non-parametric approximation of the data density for each class.

Note that the classifier used to obtain a log ratio of probabilities is a potentially important hyperparameter, and we consider both logistic regression (LogR) and linear discriminant analysis (LDA).

\begin{algorithm}[t]
\DontPrintSemicolon
\KwIn{
Trained model $f(\cdot)$, 
Pos. and Neg. Test Data $\mathcal{X}^+, \mathcal{X}^-$,
Iterations $T$,
Symbols per query $K$,
Attempts per symbol $N$,
Alphabet size $A$,
Decision threshold $\delta$,
}
\KwOut{ITR}
$C \gets 0$ \tcp*[r]{correct count}
\For(\tcp*[f]{target symbols}){$t \gets 1:T$}{
    $\pi_0 \gets (\frac{1}{A}, \ldots, \frac{1}{A})$ \tcp*[r]{unif symbol prior}
    \For(\tcp*[f]{chances to update}){$n \gets 1:N$}{
        \tcp{sample query symbols}
        $\{ q_i \}_{i\shorteq 1}^K \sim \pi_{n-1}$ \;
        \tcp{sample matching data}
        $\{ x_i \sim \mathcal{X}^+ \ \textbf{if} \ q_i = t \ \textbf{else} \ x_i \sim \mathcal{X}^- \}_{i \shorteq 1}^K$ \;
        $L \gets f(\{x_i, q_i\})$ \tcp*[r]{model likelihoods} \label{line:predict-likelihoods}
        Calc. $\pi_{n}$ from $\pi_{n-1}$ and $L$\! \tcp*[r]{Eq.\ref{eqn:unnormalized-update-seen}-\ref{eqn:normalized-update}}
        \tcp{see if target was typed}
        ind, val $\gets \argmax (\pi_n), \max(\pi_n)$ \;
        \textbf{if} \ ind$\shorteq t$ and val $\geq \delta$ \ \textbf{then} \ $C \gets C + 1$ and \textbf{break}\;
    }
}
\Return ITR($A, C/T$) \;
\caption{Estimating ITR via simulated typing. Note that the likelihood $L$ predicted by the model at each step can be either $p(e | \ell)$ or $p(\ell| e) / p(\ell)$, as described in Sec.~\ref{sec:recursive-update}.}
\label{alg:typing}
\end{algorithm}

\subsection{Single Trial Performance Metrics}

We evaluate the performance of each model in the context of single EEG trials, as well as in a simulated typing task. 
Note that the RSVP typing task inherently involves working with imbalanced classes, since the user is only interested in $1$ symbol at a time from a large alphabet. 
We therefore evaluate the performance on unseen single EEG trials by computing balanced accuracy, which is defined as the average of the accuracies for each class.

Computing balanced accuracy for discriminative models is simple, as the model directly outputs label probabilities $p(\ell | e)$. However in order to compute balanced accuracy for the generative models we compare to, we must convert their output $p(e | \ell)$ using Bayes' rule:
\begin{align}
    p(\ell | e) & = \frac{p(e | \ell) p(\ell)}{p(e)}
\end{align}
We can ignore the normalizing constant $p(e)$ as before and normalize the distribution over the alphabet. However, we must choose a prior over labels $p(\ell)$. We consider two choices of prior; an empirical prior (``Emp Prior'') based on the label fractions in the training set, and a uniform 50/50 prior (``Unif Prior'').

\subsection{Simulated Typing Procedure}

We also evaluate the performance of each model in a simulated typing task. We begin by pretraining each model on $80\%$ of the dataset, and then apply the simulated typing procedure described in Alg.~\ref{alg:typing}. The model makes $T$ independent attempts to type a letter. Each attempt consists of up to $N$ rounds of query, response, and probabilistic update as described in Sec.~\ref{sec:recursive-update}. If a letter passes the decision threshold, it is typed and may be correct or incorrect. If $N$ rounds pass without typing a letter, it is considered incorrect.

\subsection{Performance Metrics}
\label{sec:metrics}

We can characterize the performance of each model by computing its information transfer rate (ITR)~\cite{shannon1948mathematical}. Given a communication channel that is used to convey $1$ out of $A$ possible symbols at a time, and given that the fraction of correctly conveyed symbols is $P$, we have:
\begin{align}
    \text{ITR}(A, P) \coloneqq \log_2(A)+ P \log_2 P + (1 \shortminus P) \log_2 \frac{1 \shortminus P}{A \shortminus 1} \, \cdot
\end{align}
Here, ITR is measured in bits per attempted symbol. Note that in Alg.~\ref{alg:typing}, a model is given $N$ chances to query letters and update estimated probabilities; it may type nothing, or may type one symbol. For computing ITR, we do not distinguish between typing an incorrect typed symbol and failing to type any symbol. 
In this work, we are interested only in the \textit{relative} effect of methods on ITR, rather than the absolute performance of a typing system. For this reason, we do not convert estimated ITR into bits per time, as this depends on additional factors that we hold fixed between models, such as the delay between stimuli, and the delay between queries.

\section{Results}

\begin{table}[t!]
\caption{Balanced Accuracy and Information Transfer Rate (ITR) for Discriminative (Disc) and Generative (Gen) Models. The discriminative strategy yield models with higher balanced accuracy and information transfer rates. Entries show mean and standard deviation across $5$ random train/test splits. Control models use the discriminative strategy but always assign high probability to a fixed class. See Sec.~\ref{sec:metrics} for ITR calculation.}
\label{tab:results}
\centering
\begin{tabular}{llll}
\bottomrule
Strategy & Model              & Balanced Acc   & ITR            \\ 
\toprule	
\bottomrule
Disc & LogR & 0.730 ± 0.001 & 0.817 ± 0.047 \\ 
Disc & EEGNet & 0.745 ± 0.003 & 0.930 ± 0.050 \\ 
Disc & 1D CNN & \textbf{0.782 ± 0.005} & \textbf{1.103 ± 0.047} \\ 
Disc & 2D CNN & \textbf{0.779 ± 0.004} & \textbf{1.153 ± 0.068} \\ 
\toprule	
\bottomrule
Gen & LDA (Emp Prior) & 0.509 ± 0.000 & \multirow{2}{*}{0.678 ± 0.077} \\ 
Gen & LDA (Unif Prior) & 0.687 ± 0.003 &  \\ \hline
Gen & LogR (Emp Prior) & 0.500 ± 0.000 & \multirow{2}{*}{0.218 ± 0.022} \\ 
Gen & LogR (Unif Prior) & 0.694 ± 0.002 &  \\ \toprule	
\bottomrule
Control & Always Class 0 & 0.500 ± 0.000 & 0.000 ± 0.000 \\ 
Control & Always Class 1 & 0.500 ± 0.000 & 0.000 ± 0.000 \\ 
\toprule
\end{tabular}
\end{table}

\begin{figure}[htb]
    \centering
    \includegraphics[width=0.9\columnwidth]{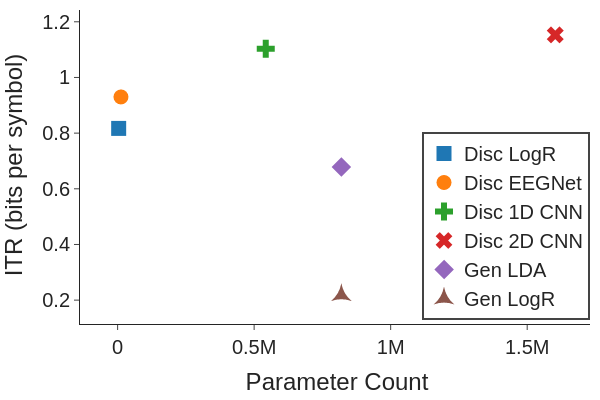}
    \caption{Information Transfer Rate vs Model Size. Discriminative (Disc) models outperform generative (Gen) models across a wide range of sizes. Among Disc models, performance increases with model size.}
    \label{fig:itr-vs-params}
\end{figure}

We find that the proposed discriminative modeling approach outperforms the previous generative modeling strategies in terms of balanced accuracy and ITR on our simulated typing task. In Table~\ref{tab:results}, we see that the overall best performing models use a neural network classifier. In the balanced accuracies of generative models, we see that using an empirical prior results in performance close to random chance, while using a uniform prior improves the accuracy - this is due to the highly imbalanced classes in our dataset.

We also find that the proposed discriminative modeling strategy performs well for both large and small models. In Fig.~\ref{fig:itr-vs-params}, we show model ITR as a function of parameter count. Discriminative models outperform generative models across a wide range of model sizes. Among the discriminative models, there is also a positive trend of model performance with increasing model size, indicating that increasing model size may provide additional benefit.

\section{Conclusion}

In this work we focus on accurate posterior estimation of user intent from EEG during an RSVP typing task. 
We build upon a variety of recent work performing Bayesian updates to symbol probabilities that largely use generative models for computing updates.
We formulate the RSVP typing task using a probabilistic graphical model and derive a novel approach for recursive Bayesian updates in this context that relies only on computing discriminative probabilities.
This allows us to make use of deep neural network classifiers for estimating these probabilities.
We find that these neural network classifier models achieve much higher balanced accuracy than comparable generative modeling approaches, for several choices of prior.
We construct a simulated typing task to mimic the real RSVP setting, so that we can estimate ITR for various models. 
We find that our proposed method outperforms previous generative modeling strategies, achieving much higher ITR.

Our results demonstrate that the proposed method has the potential to improve communication speeds in BCIs in the query-and-response paradigm. 
The same approach used here may benefit from more powerful classification models, and may be applicable in the setting of single-subject training by taking advantage of few-shot and transfer learning algorithms for classifier training. 

\bibliographystyle{IEEEtran}
\bibliography{references}

\end{document}